\documentclass[a4paper,12pt,nofootinbib,notitlepage,onecolumn,amsmath,amssymb,aps,prd,longbibliography]{revtex4-1}
\usepackage[]{amsmath}

\usepackage{hyperref}
\usepackage{braket}
\usepackage{overpic}
\usepackage[left=3cm,right=3cm]{geometry}

\def\prd{Phys. Rev. D} 
\def\prl{Phys. Rev. Lett.} 
\def\mnras{Mon. Not. R. Astron. Soc.}

\begin{document} 


\title{Dragging of inertial frames by matter and waves${}^1$}

\author{JI\v{R}\'{I} BI\v{C}\'{A}K}
\author{TOM\'{A}\v{S} LEDVINKA}

\address{Institute of Theoretical Physics, Charles University
\\
V Hole\v{s}ovi\v{c}k\'{a}ch 2, Prague, Czech Republic}


\begin{abstract} 
In this paper, we review and analyze four specific general-relativistic problems in which gravitomagnetism plays an important role: the dragging of magnetic fields around rotating black holes, dragging inside a collapsing slowly rotating spherical shell of dust, compared with the dragging by rotating gravitational waves. 
We demonstrate how the quantum detection of inertial frame dragging can be accomplished by using the Unruh-DeWitt detectors. Finally, we shall briefly show how ``instantaneous Machian gauges'' can be useful in the cosmological perturbation theory.
\end{abstract}



\maketitle

\section{Introduction}
\footnotetext[1]{{Based on a plenary talk presented at the Sixteenth Marcel Grossmann Meeting on Recent Developments in Theoretical and Experimental General Relativity, Astrophysics and Relativistic Field
Theories, online, July 2021.}}

The relativistic effect of the dragging of inertial frames is associated with the profound criticism of the Newtonian concepts of absolute space and time by Mach. It was this criticism which appears to be one of the most influential for Einstein in the creation of general relativity. 
Ernst Mach (1838-1916) was born very close to Brno (like Kurt Goedel), today Czech Republic, formerly Austria-Hungary. He was professor at the Karl–Ferdinands Universt\"at in Prague for 28 years and his influence on Prague physics indirectly also led to the stay of Albert Einstein in Prague. 
    
Let us illustrate Mach's ideas by just one thought from his most influential work:
\begin{quote}
 [The] investigator ... must feel the need of ... knowledge of the immediate connections, say, of the masses of the universe. There will hover before him as an ideal insight into the principles of the whole matter, from which accelerated and inertial motions will result in the same way \emph{(Science of mechanics)}.
 \end{quote}
Keeping the Mach tradition, on the 150th anniversary of the birth of Ernst Mach, the international conference was organized in September 1988 at the Charles University in Prague. The conference papers are published in the book \cite{Prosser91}, including several contributions by some leading scientists, philosophers and historians of science. The meeting was inspiring for Julian Barbour and Herbert Pfister who, during the meeting, decided to organize the meeting on just Mach's principle. A comprehensive volume \cite{bucket95}, based on the conference at T\"{u}bingen in July 1993, includes not only many contributions by leading experts but also detailed texts recording many discussions.

Prague and Brno historically became attractive places for a number of influential scientists. Here we naturally recall the stay of Albert Einstein at the Karl–Ferdinands Universt\"at from April 1911 until July 1912. His invitation to Prague was strongly supported by Mach's Prague pupils. Einstein wrote several pioneering papers showing the route to the final version of General Relativity, in particular in his answer to Max Abraham how a future theory of gravity should look like. In the best known paper from Prague, he forecasted the light bending (we refer, e.g., to \cite{Bicak2012AEPRG} for details). From the point of view of dragging, however, it is most interesting that it was in Prague where this phenomenon was first discussed, albeit based on Einstein's Prague preliminary version of general relativity. In his work \cite{AE1912} he considers 
a shell of matter and its influence on a mass point placed in its center as the shell starts to accelerate.
In his words:
\emph{This suggests that the entire inertia of a mass point is an effect of the presence of all other masses, which is based on a kind of interaction with the latter (this is exactly the same point of view that E. Mach advanced in his astute investigations on this subject).}

\subsection*{Experiments}

\begin{figure}
\begin{center}
\includegraphics[width=60mm]{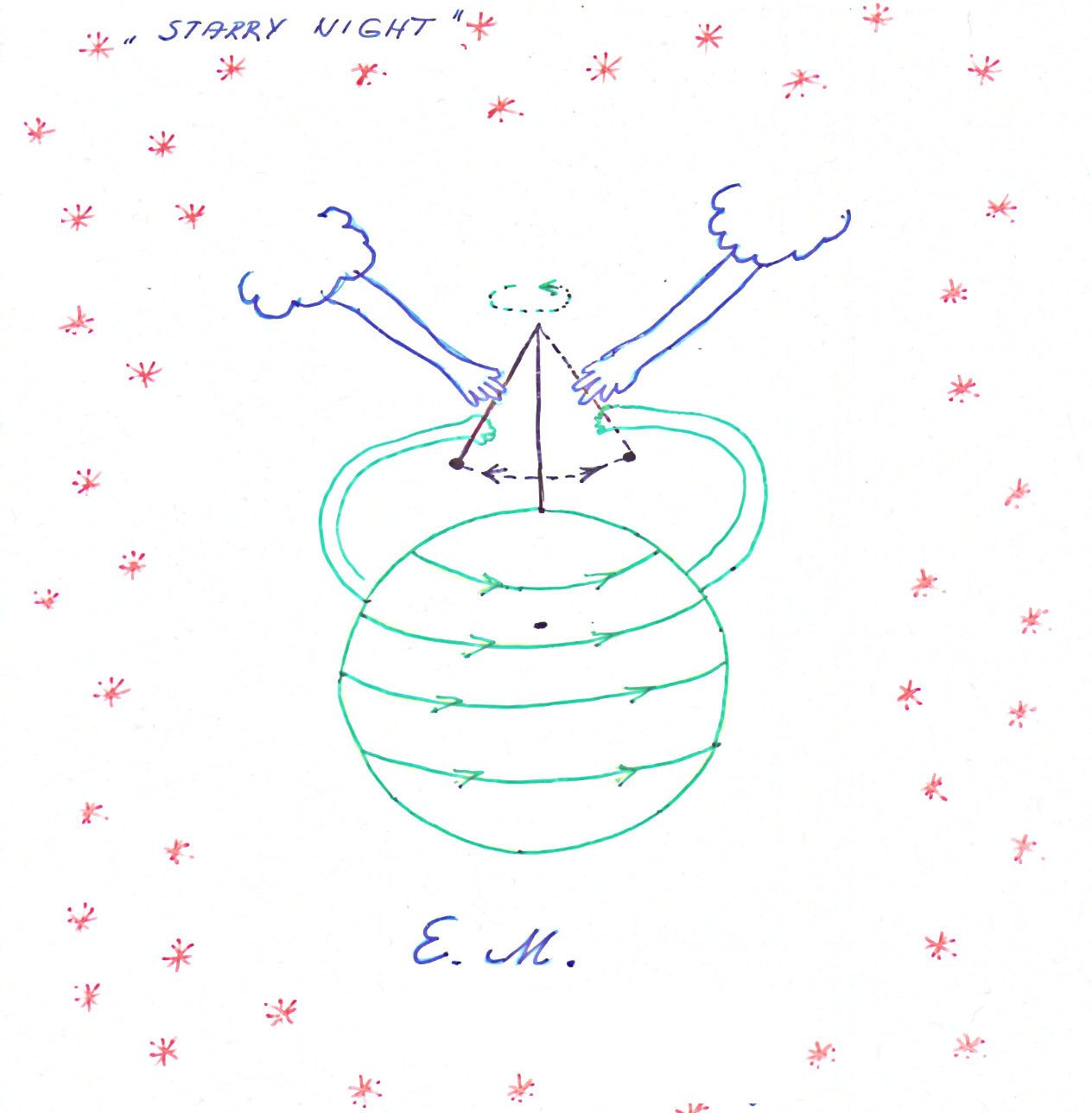}
\includegraphics[width=65mm]{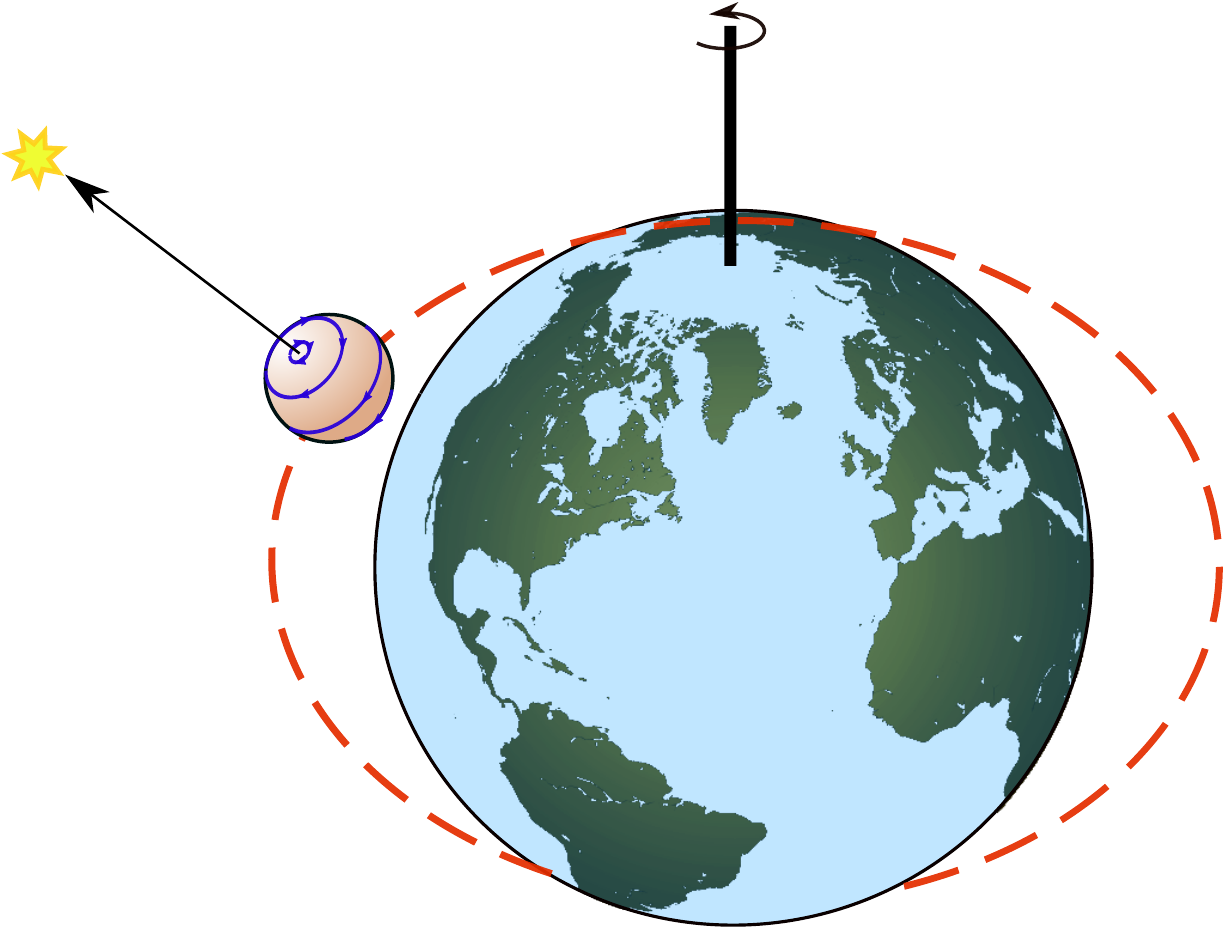}
\end{center}
\caption{\emph{Left:} Dragging of the pendulum plane observed at the pole seen as a Machian competition between masses of distant stars and Earth. 
\emph{Right:} A gyroscope on the polar orbit. Its axis changes direction both due 
to the geodesic precision ($\approx 6 600  \text{mas y}^{-1}$) and due to frame dragging induced by Earth rotation ($\approx 39  \text{mas y}^{-1}$). This change with respect to a distant ``guide'' star was measured by the Gravity Probe B space experiment.}
\label{fig:hands}
\end{figure}

A nice experiment to measure rotational dragging was suggested by Braginsky, Polnarev and Thorne in 1984 \cite{Braginskii1984}. The plane of a Foucault pendulum
at the South Pole will be fixed with respect to ``fixed stars'' around which on average will not produce  the dragging of the pendulum into the rotation. However, slowly rotating ``very close'' Earth does produce the effect of
$\omega_\text{drag} = 2 J/R^3$, where $J$ and $R$ are angular momentum and radius of the Earth --- see Fig. \ref{fig:hands} left.  

Most sophisticated experiment to confirm the dragging of inertial frames by the rotating Earth is, of course, Gravity Probe B. 


The idea of placing a gyroscope on a free orbit around the Earth was conceived independently by Schiff and Pugh. In fact, the gyros (the ``most spherical balls'' produced by man) were four. There was also a telescope in the satellite with the gyros which was pointing towards the Guide Star---see Fig. \ref{fig:hands}. The launch occurred on April 20, 2004, and lasted 16 months. The first results appeared in April 2007 but the complete analysis was finished only in 2015 (see \cite{Everitt_2015}). The measured frame dragging effect, $-37.2 \pm 7.2\; \text{mas y}^{-1}$,
confirmed the general-relativistic prediction $-39.2 \pm 0.2\; \text{mas y}^{-1}$.  
The relatively large error was caused primarily by random patches of electric charge on rotors (gyros) and their housing.

A very nice experiment demonstrating the dragging effects on the nodal rates of 3 laser-ranged satellites using the Earth gravity field model produced by space mission GRACE was performed by the group of I. Ciufolini (see his plenary talk at this conference, contribution by Lucchesi in the Session PT5; see also the book on Gravitomagnetism by Ciufolini and Wheeler \cite{Ciufolini1995}).

\section{Magnetic fields, Meissner effect and dragging} 
Consider first the magnetic test field $B_0$ which is uniform at infinity and aligned with the hole rotation axis. Solution of Maxwell’s equations on the background geometry of a rotating (Kerr) black hole with boundary condition of uniformity at infinity and finiteness at the horizon yields the field components; from these the lines of force are defined as lines tangent to the Lorentz force experienced by test magnetic/electric charges at rest with respect to locally nonrotating frames (preferred by the Kerr background field). The field lines are plotted in Fig. 2 for $a = 0.5M$ and in the extreme case when $a = M$. Note that only weak expulsion occurs in the former case. There is a simple analytic formula for the flux across the hemisphere of the horizon, see \cite{King1975,BiJa,BiLe2000}:
\begin{equation}
\Phi = B_0 \pi r^2_+\left(1 - \frac{a^4}{r^4_+}\right),
\end{equation}
where $r_+ = M + (M^2 - a^2)^{1/2}$.
As a consequence of the coupling of magnetic field to frame-dragging effects of the Kerr geometry the electric field of a quadrupolar nature arises. Its field lines are shown in Fig. \ref{fig:KerrInducedE}. Again the field lines are expelled: while even with $a = 0.95M$ it is still not very distinct, the expulsion becomes complete in the extreme case. One can demonstrate that total flux expansion takes place for all axisymmetric stationary fields around a rotating black hole \cite{BiJa,BiLe2000}. In Fig. \ref{fig:MeissnerLoop} the field lines of a current loop in the equatorial plane are shown. The Meissner-type effect arises also for charged (Reissner-Nordstr\"{o}m) black holes as shown in Fig. \ref{fig:MeissnerLoop} right.

Although extremely charged black holes ($e^2 = M^2$) are probably not important in astrophysics they may be significant in fundamental physics (as, for example, very special supersymmetric BPS states mass of which does not get any quantum corrections). 
In the charged case the electromagnetic perturbations are in general coupled to gravitational perturbations, the resulting formalism is involved. Neverthe-\newline less, one may construct explicit solutions, at least in stationary cases. From these the magnetic field lines follow as in the Kerr case. The magnetic field lines of a dipole located far away from the hole look like in a flat space, however, when the dipole is close to the horizon, the expulsion in the extreme case is evident (Fig. \ref{fig:MeissnerLoop} right). Due to the coupling of perturbations closed field lines appear without any electric current inside; see \cite{BiDvo1980} for details.
There exist exact models (exact solutions of the Einstein-Maxwell equations) representing in general rotating, charged black holes immersed in an axisymmetric magnetic field. The expulsion takes place also within this exact framework - see \cite{BiKa,KaVo1991,Budinova2000}.
Recently, the Meissner effect was also demonstrated for extremal black-hole solutions in higher dimensions in string theory and Kaluza-Klein theory. The question of the flux expulsion from the horizons of extreme black holes in more general frameworks is not yet understood properly. The authors of \cite{Chamblin1998}
``believe this to be a generic phenomenon for black holes in theories with more complicated field content, although a precise specification of the dynamical situations where this effect is present seems to be out of reach.''

\begin{figure}
\begin{center}
\includegraphics[width=55mm]{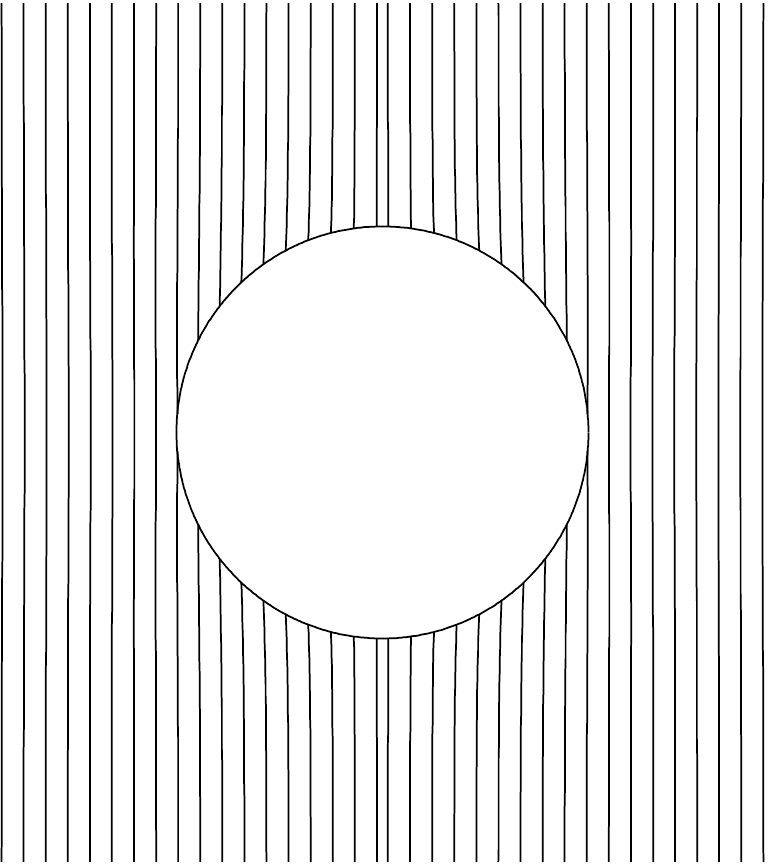}
~~~~
\includegraphics[width=55mm]{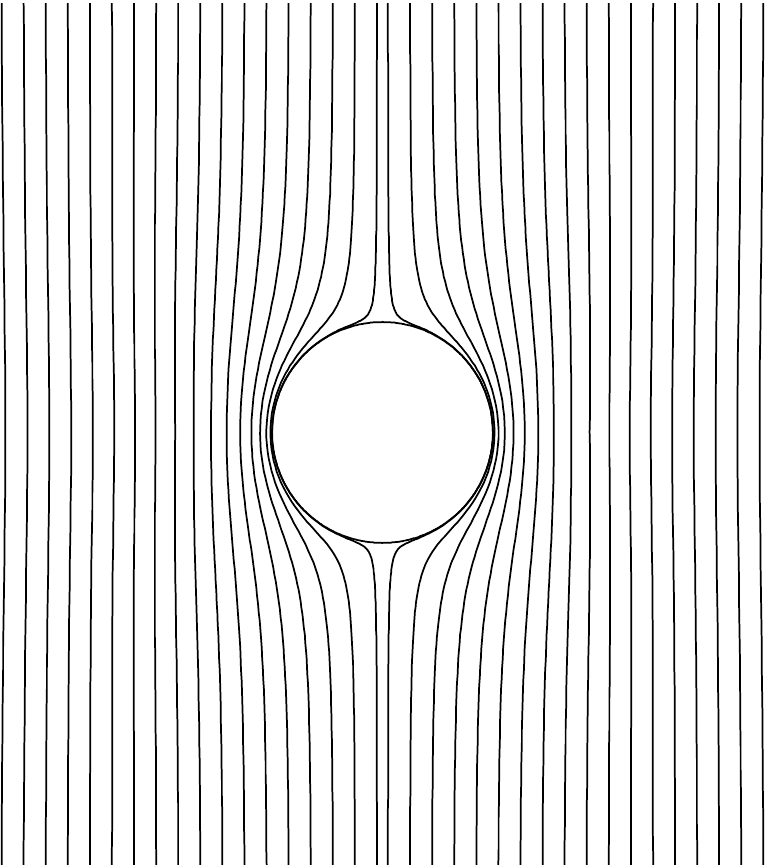}
\end{center}
\caption{Field lines of the test magnetic field uniform at infinity and aligned with the hole rotation axis. Two cases with a = $0.5M$ (left) and $a = M$ (right) are shown.}
\label{fig:KerrExpulsion}
\end{figure}

\begin{figure}
\begin{center}
\includegraphics[width=55mm]{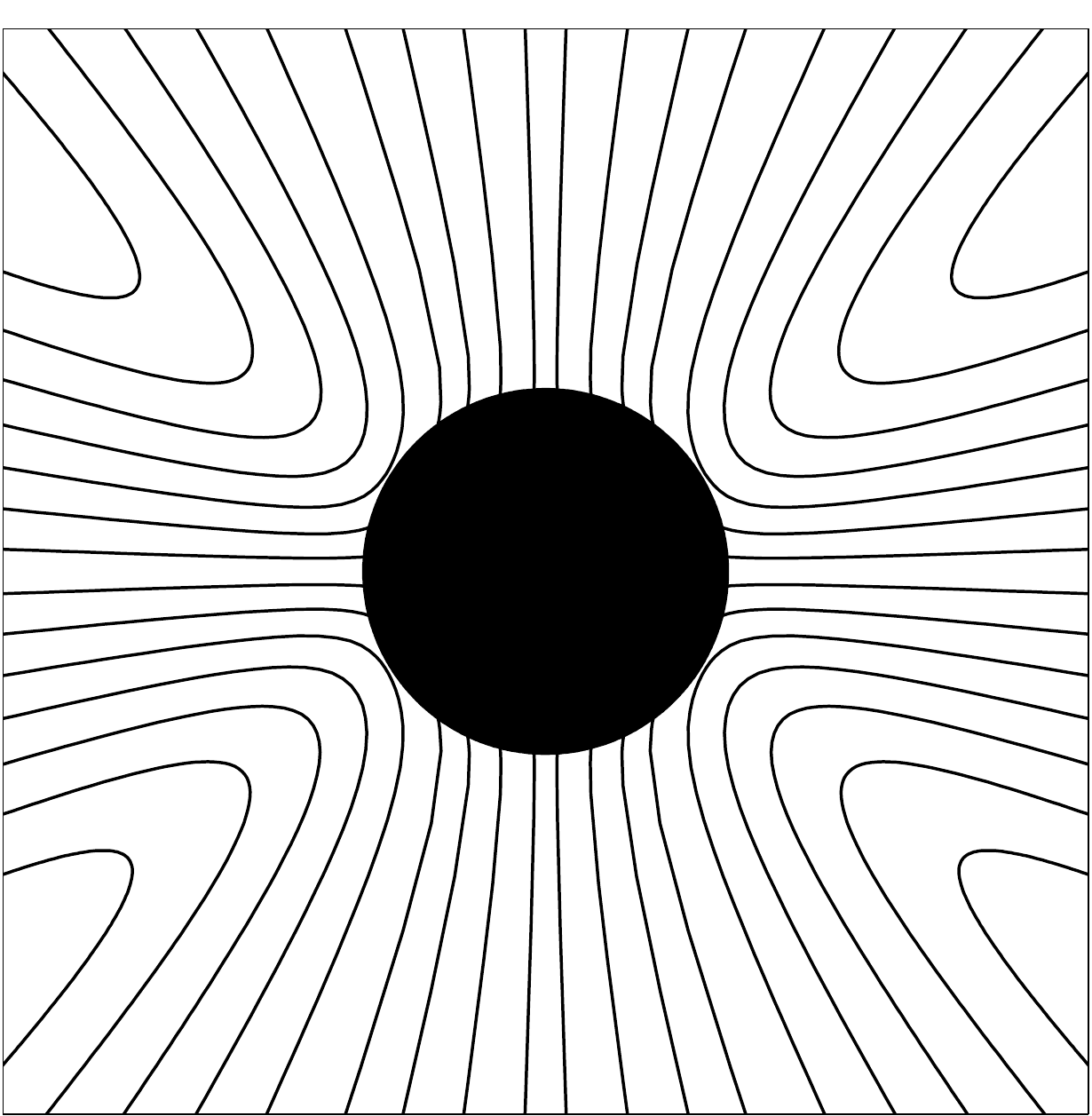}
~~~~
\includegraphics[width=55mm]{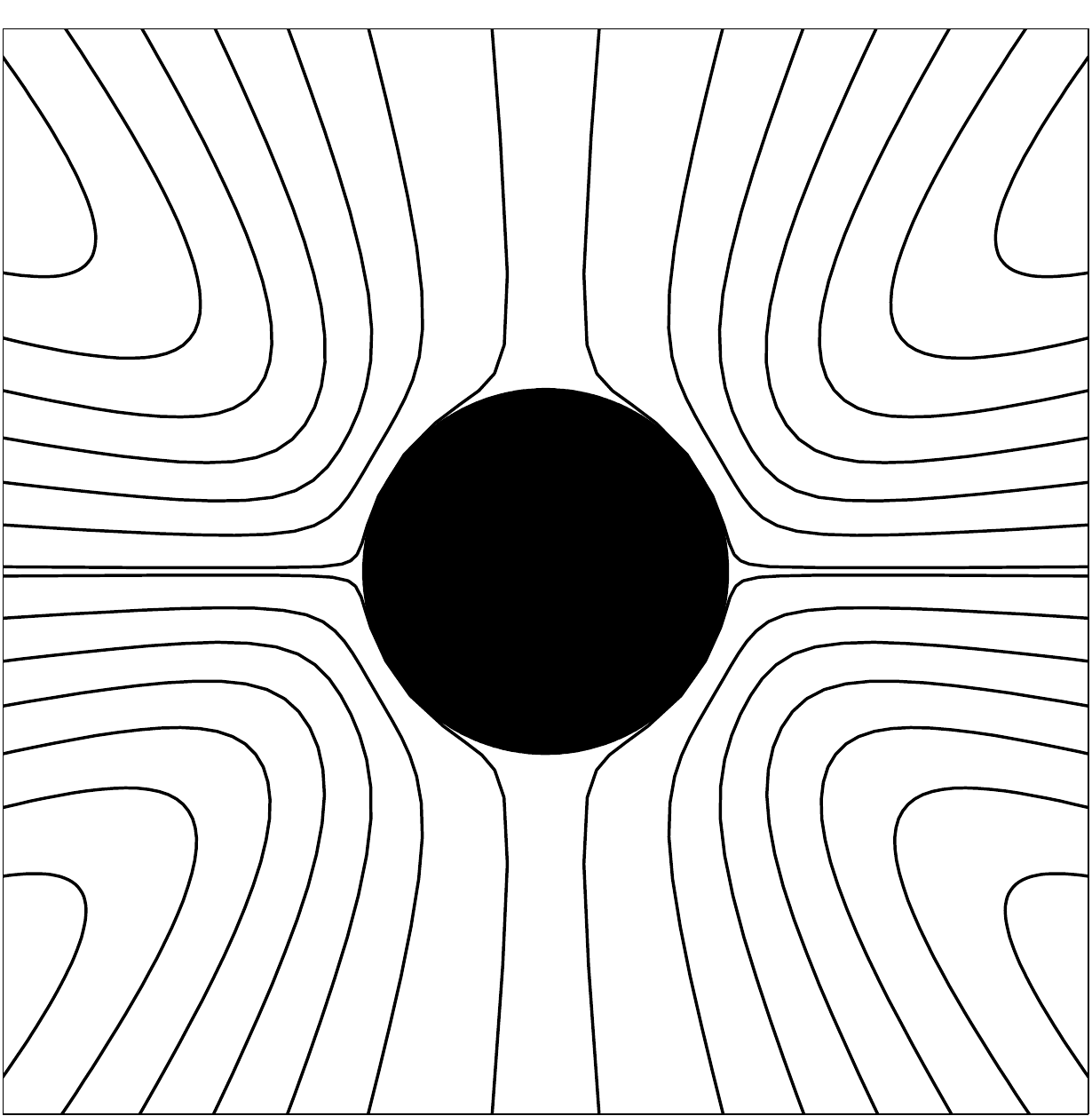}
\end{center}
\caption{Field lines of the electric field induced by the ``rotating geometry'' of Kerr black hole in asymptotically uniform test magnetic field; $a = 0.95M$ (left), and $a = M$ (right). 
}
\label{fig:KerrInducedE}
\end{figure}

\begin{figure}
\begin{center}
\includegraphics[width=55mm]{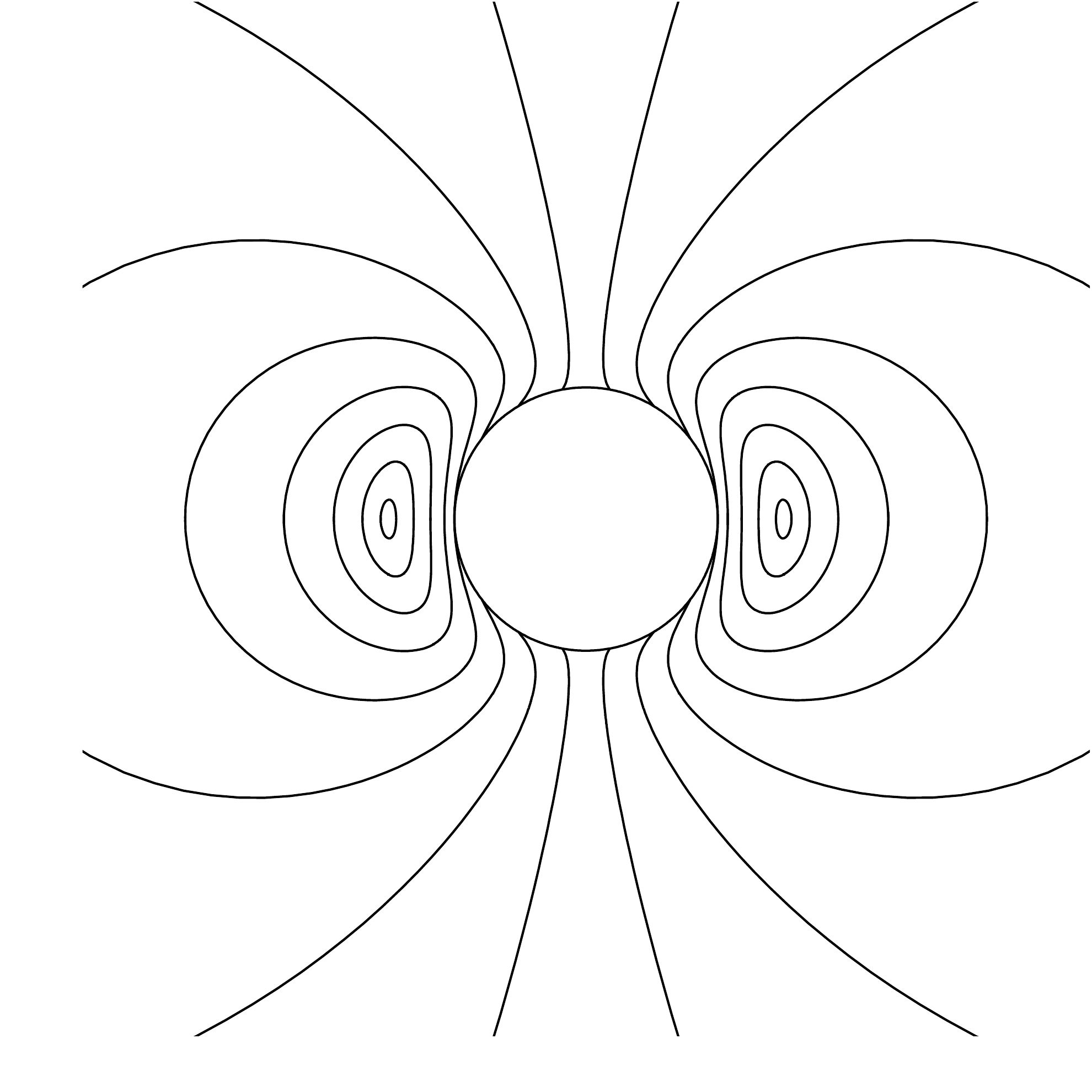}
~~~~
\includegraphics[width=55mm]{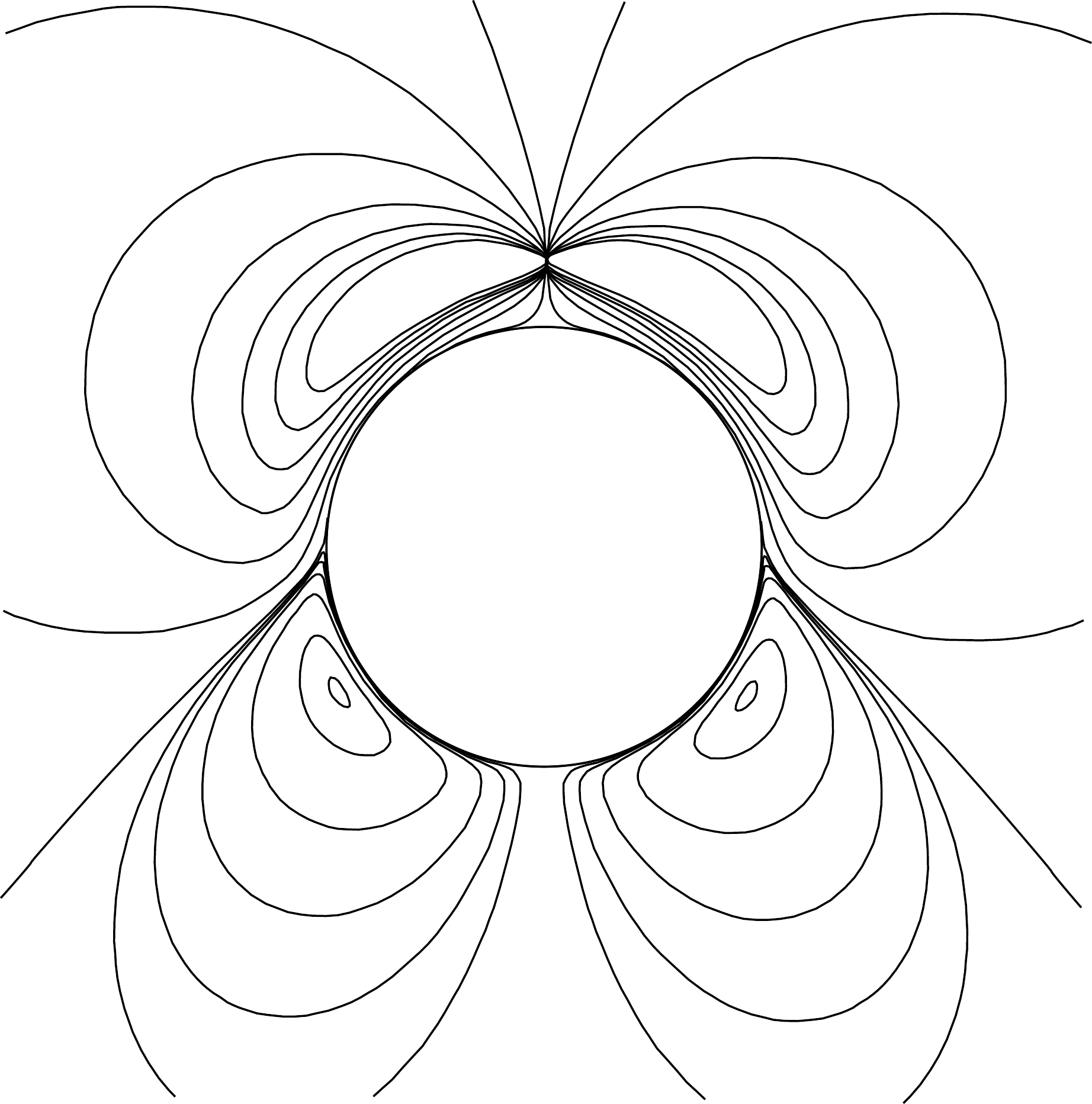}
\end{center}
\caption{\emph{Left:} Field lines of the test magnetic field of a current loop in the equatorial plane of the Kerr spacetime with $a=0.995M$.
\emph{Right:} Field lines of the test magnetic field of a magnetic dipole 
placed near the extreme Reissner-Nordstr\"{o}m black hole.
}
\label{fig:MeissnerLoop}
\end{figure}

The flux expulsion does not occur when the configuration is not axially symmetric.
The electromagnetic field occurring when a Kerr black hole is placed in an originally uniform magnetic field without assuming  the alignment of the direction of the magnetic field and the axis of symmetry of the black hole was first given in \cite{BiDvo} (see also \cite{PolB}, \cite{BiJa}).

\begin{figure}
\begin{center}
\includegraphics[width=9cm]{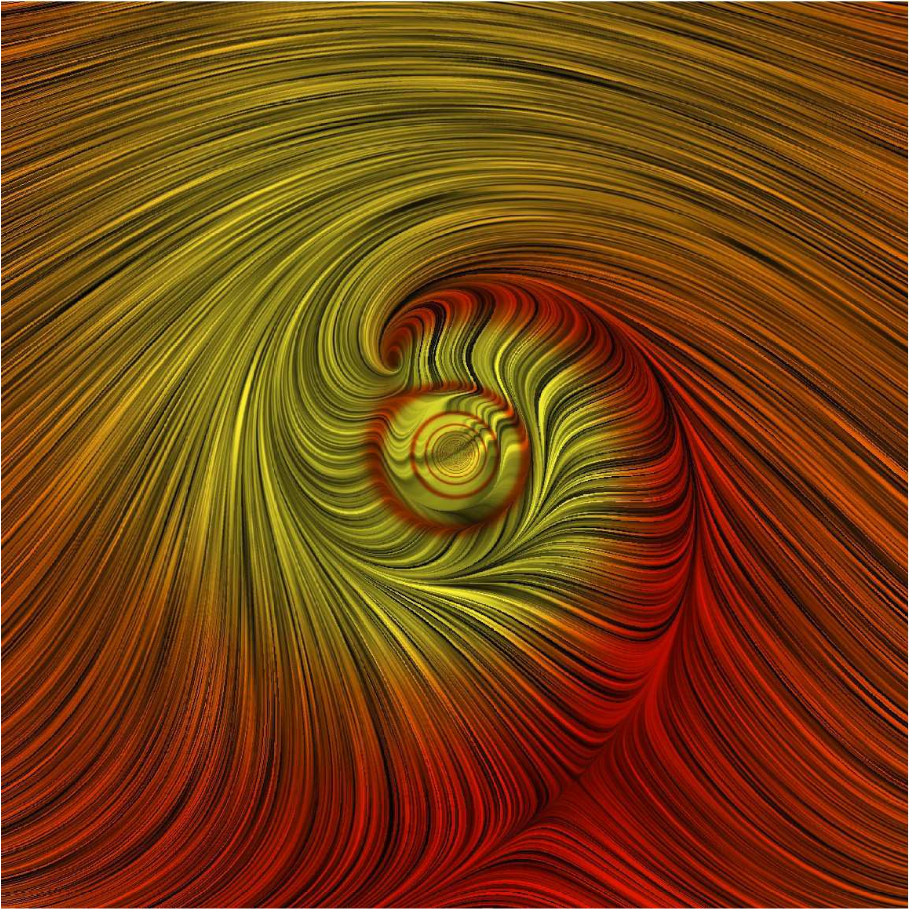}
\end{center}
\caption{
Field lines in the equatorial plane of the Kerr black hole with
color indicating the intensity of the field. 
Field lines which are asymptotically uniform and perpendicular to the rotation axis are dragged by rotation in vacuum (no conductive medium around).
The horizon is in the center as a point. The critical point appears where the field vanishes (approx. at 11 hours).
The figure is taken from \cite{KKK}.
}
\label{fig:KKKfieldlines}
\end{figure}

The properties of these ``oblique'' fields and their possible astrophysical relevance were already studied in the contribution \cite{BiKa} to the 5th Marcel Grossmann Meeting in Perth. They were then much developed in a number of important papers by Karas and his group appearing until today.  Here we mention just few results and refer to the paper by Karas given in the Session PT5 of this MG16 meeting. 
One of the effects of the rotation on the fields which are asymptotically uniform and perpendicular to the rotation axis is the dragging of field lines by rotation and, as a consequence, the appearance of critical points where the field vanishes as seen in Fig. \ref{fig:KKKfieldlines}.

In the most recent work, Karas \emph{et al.} realized that due to the presence of the plasma in the accretion flows and differential rotation even weak electromagnetic fields are crucial. Although magnetic fields within the accretion flow are turbulent in almost empty funnels around the rotation axis they can be organized on large scales and it is from here where they can accelerate the charged particles and produce collimated jets. Most recently, Karas and 
Kop\'{a}\v{c}ek conclude that inclined field (its oblique component) leads to more efficient acceleration and larger final Lorentz factors of escaping particles; see \cite{KaKo}
and number of references therein.
For a leading expert view on formation of jets and black-hole shadow in case of M87, see the contribution of R. Blandford to the Session PT5 of this MG16 meeting.

\section{Dragging by a slowly rotating, collapsing spherical shell}

\begin{figure}
\begin{center}
\includegraphics[width=41mm]{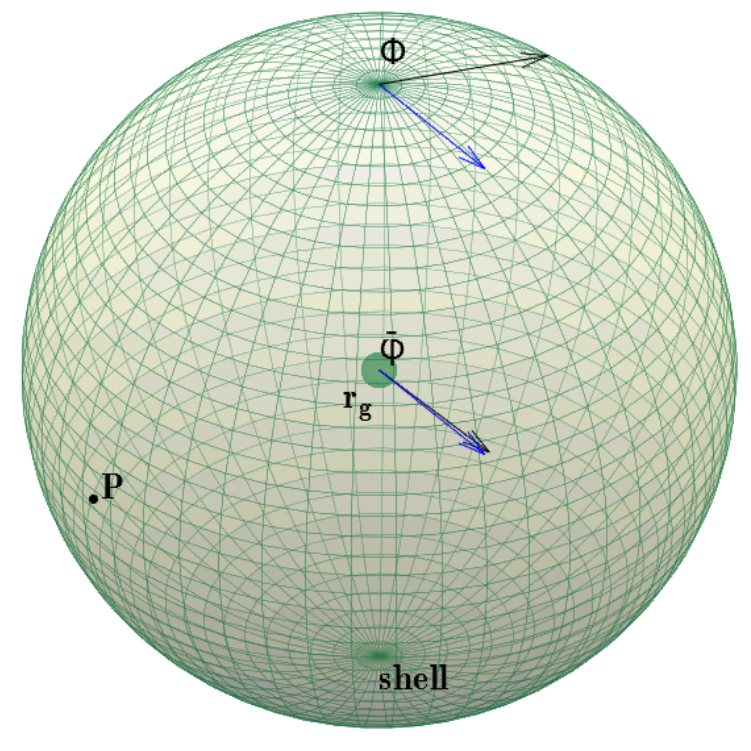}
\includegraphics[width=41mm]{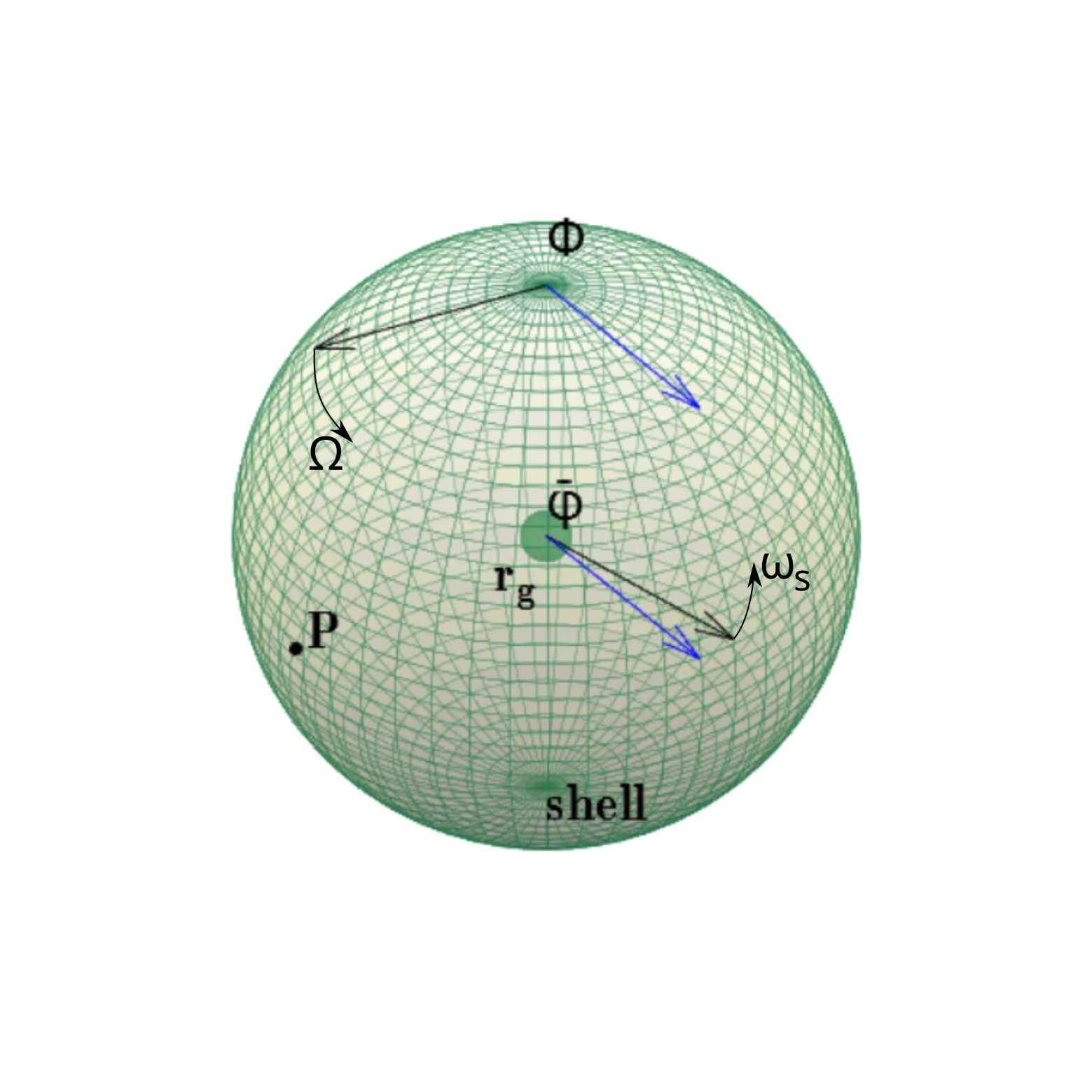}
\includegraphics[width=41mm]{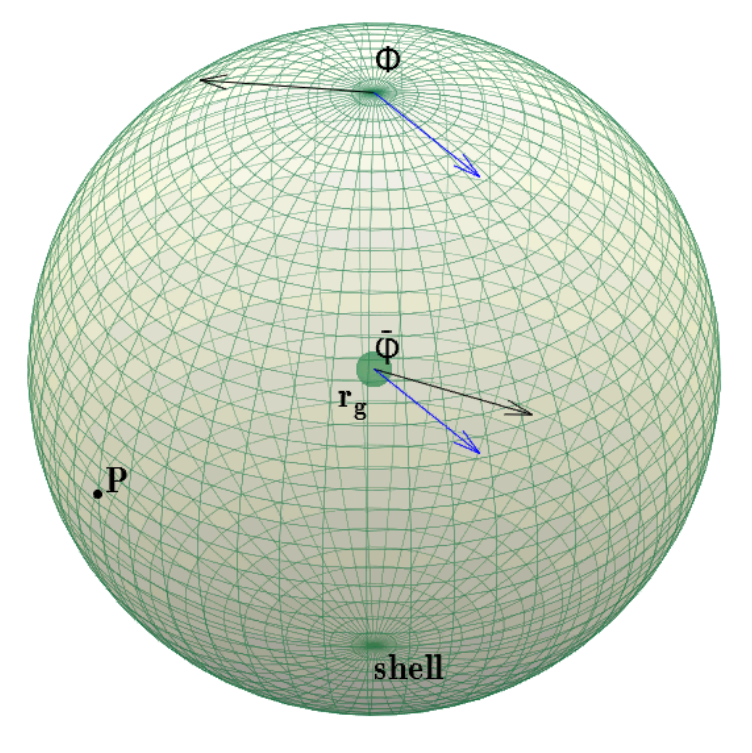}
\end{center}
\caption{
Dragging of a gyroscope inside a slowly rotating, collapsing and rebouncing thin shell.
Angular velocities of the vectors indicating the shell rotation $\Omega$ and the central inertial frame  rotation
$\omega_s$ are shown in the middle panel. See \cite{Barker2017CQG} for details (stills from the animation  \url{https://utf.mff.cuni.cz/~ledvinka/psi/a1.mp4}  
 by W. Barker.)}
\label{fig:BarkersSpheres}
\end{figure}

A spherical shell in slow rotation and collapse  (see Fig. \ref{fig:BarkersSpheres}) produces slightly perturbed Schwarzschild spacetime outside with the metric
\begin{equation}
 ds^2 \doteq \left(1-\frac{2M}{r}\right)\,dt^2
 -\left(1-\frac{2M}{r}\right)^{-1}dr^2-r^2\,d\theta^2-r^2
 \left(d\phi-\omega dt\right)^2,
\end{equation}
where $\omega$ is the frame dragging potential given by $\omega=2J/r^3$, 
$J$ is fixed (small) total angular momentum of the shell. At the shell’s surface
$r=r_s$ is decreasing as shell collapses and $\omega_s = 2J/r_s^3$ is increasing.
Notice (Fig. \ref{fig:BarkersSpheres}) that $\Omega = d \phi_s/dt$ is the angular velocity of the shell, $r_s^2 \Omega^2$ is neglected.

The spacetime inside the shell is flat in this approximation. Its metric $ds^2=dt^2-dr^2-r^2\;d\theta^2-r^2\sin^2\theta\;d{\bar\phi}^2$ can be joined across the shell to the metric outside.
Because $d{\bar\phi} = d\phi-\omega_s\,dt$, the local \emph{inertial} frames (LIFs) 
inside ($\bar\phi=\text{const.}$) 
all rotate rigidly with the same angular velocity   with respect to the observers at rest relative to infinity 
(``static observers'' with $\phi=\text{const}$). 
Thus ${d{\bar\phi}}/{dt}=0$ implies the time-dependent angular velocity
${d{\phi}}/{dt}=\omega_s(t)$ of the rigid rotation.

As measured \emph{in LIF's} own proper time the rate of rotation
is 
\begin{equation}
\frac{d{\phi}}{d\bar t}=\bar\omega_s = \left.\omega_s\frac{dt}{d\bar t}\right|_s .
\end{equation}
Static observers inside experience \emph{Euler acceleration}
(Coriolis$\,\sim\!\omega_s^2$, centrifugal$\,\sim\!\omega_s^2$)
and the congruence of their \emph{worldlines twists}.
Rate of rotation $\bar \Omega_\tau$ of the shell itself
measured in its proper comoving time $\tau$ is
\begin{equation}
 \bar \Omega_\tau =  \frac{3r_s}{4m_s}\omega_s  =\frac{3J}{2m_s r_s^2}.
\end{equation}

Many details about this system can be found in Refs. \cite{Katz1998CQG,Barker2017CQG,Pfister2015}.

\section{Quantum detection of inertial frames dragging}

Recently, we studied quantum Unruh-DeWitt detectors \cite{dW} and their suitability, at least in principle, for the detection of the dragging of inertial frames \cite{CBKM1} and for the detection of a conicity of space \cite{CBKM2}. We have shown, for the first time as far as we know, that the dragging of inertial frames (as well as conicity) can be observed by a quantum detector. We studied the response function of UdW detector placed in a slowly rotating shell which has flat spacetime inside and slowly rotating Kerr metric outside, as discussed in Sec. 3. (Here we assume the shell to be stationary, not collapsing.)  

The detector is a two-state system with energy gap $\Omega$ and the field-interaction Hamiltonian 
  $\hat H(\tau) = \lambda\chi(\tau)\hat\mu(\tau)\otimes\hat\Psi(x(\tau))$,
  where $\chi(\tau)$ is the switching function of the detector (ensuring that the interaction duration is $\Delta\tau = \pi/k$), $x(\tau)$ its worldline, $\hat\mu(\tau)$ its monopole momentum operator and $\lambda$ is the coupling constant.
We assume the detector-field system is in initial state 
$\Ket{0}_D \Ket{0}_\Phi$.
Then the transition probability $P$ to
$\Ket{1}_D$ is
\begin{equation}
P = \lambda^2 {\mathcal{F}}+O(\lambda^4).
\end{equation}
The response function $\mathcal{F}$ of the detector turns out to be
\begin{equation}
\mathcal{F}=
\int_{-\infty}^{\infty}  \,\int_{-\infty}^{\infty} \chi  (\tau_1)\chi(\tau_2) e^{-i\Omega(\tau_2-\tau_1)} W(x(\tau_1),x(\tau_2))\;d \tau_1\; d \tau_2,
\end{equation}
where the Wightman function of the field is
\begin{equation}
W(x(\tau_1),x(\tau_2))= {}_{\raisebox{-0pt}{${}_\Phi$}}\!\!\Bra{0}\hat\phi(x(\tau_2))\hat\phi(x(\tau_1))\Ket{0}_\Phi.
\end{equation}

\begin{figure}
\begin{center}
\includegraphics[width=85mm]{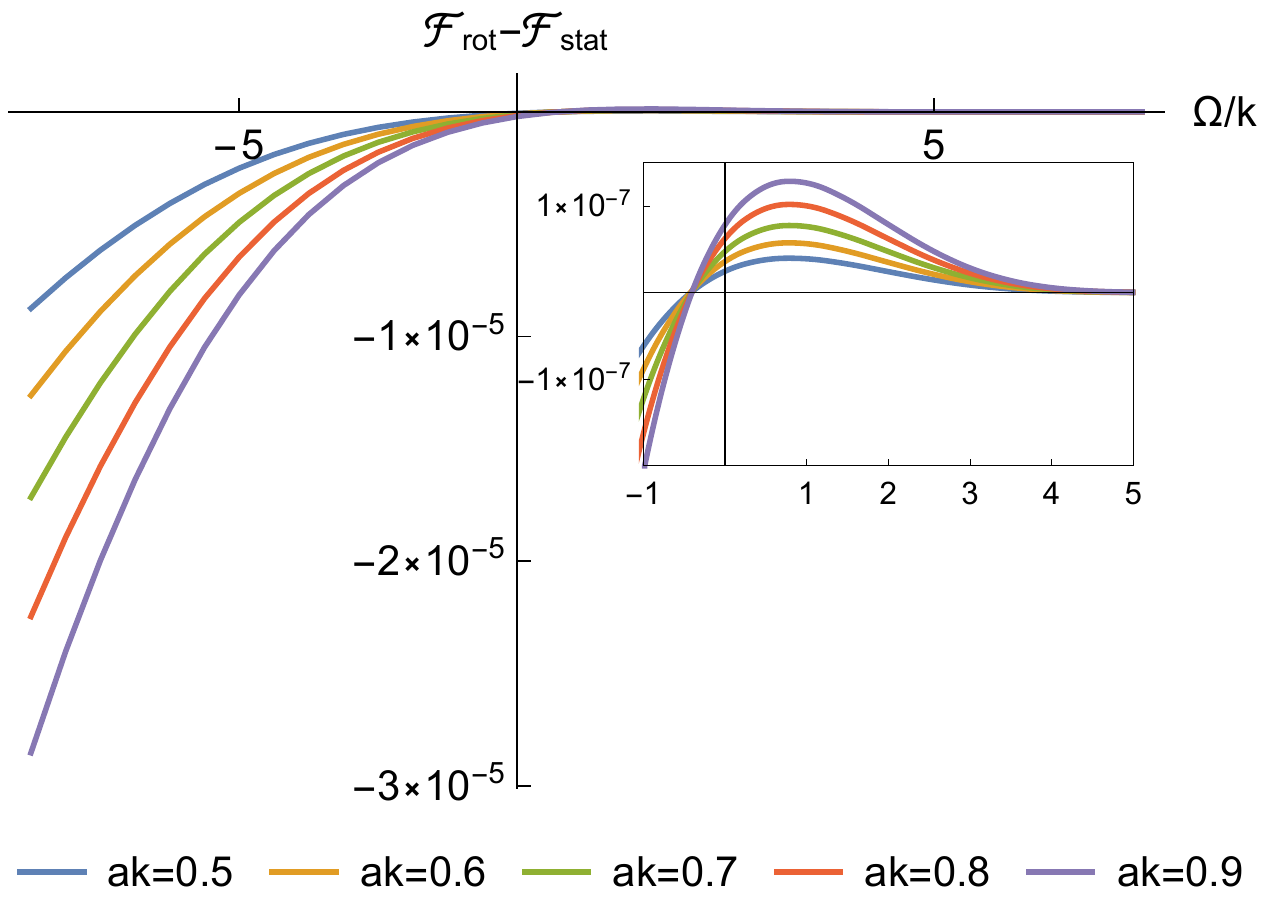}
\end{center}
\caption{
Comparison with the detector in a static shell.
Detector response function inside a slowly rotating shell is plotted for several values of the shell angular momentum $J=Ma$ appearing in the dimensionless parameter $a k$. 
The difference  $\mathcal{F}_{rot}-\mathcal{F}_{stat}$ is plotted as a function of the energy gap $\Omega$ of the detector. Shell mass $M$ and radius $R$ radius satisfy $M k=1,\, R k=3$, detector is placed at $r_d=0.5/k$ from the center. 
}
\label{fig:WanCongFig1}
\end{figure}

\begin{figure}
\begin{center}
\includegraphics[width=60mm,trim=0 1cm 0 -1cm]{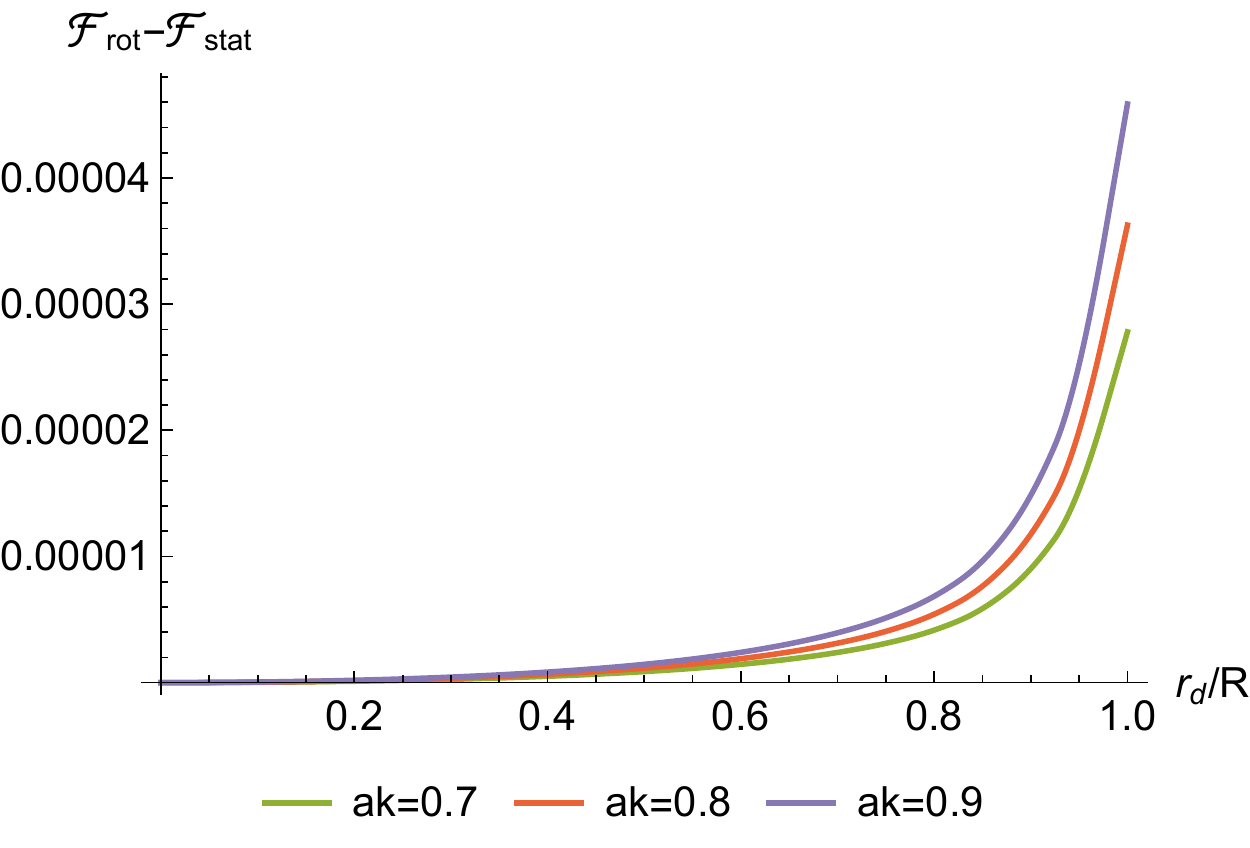}
~~
\includegraphics[width=60mm]{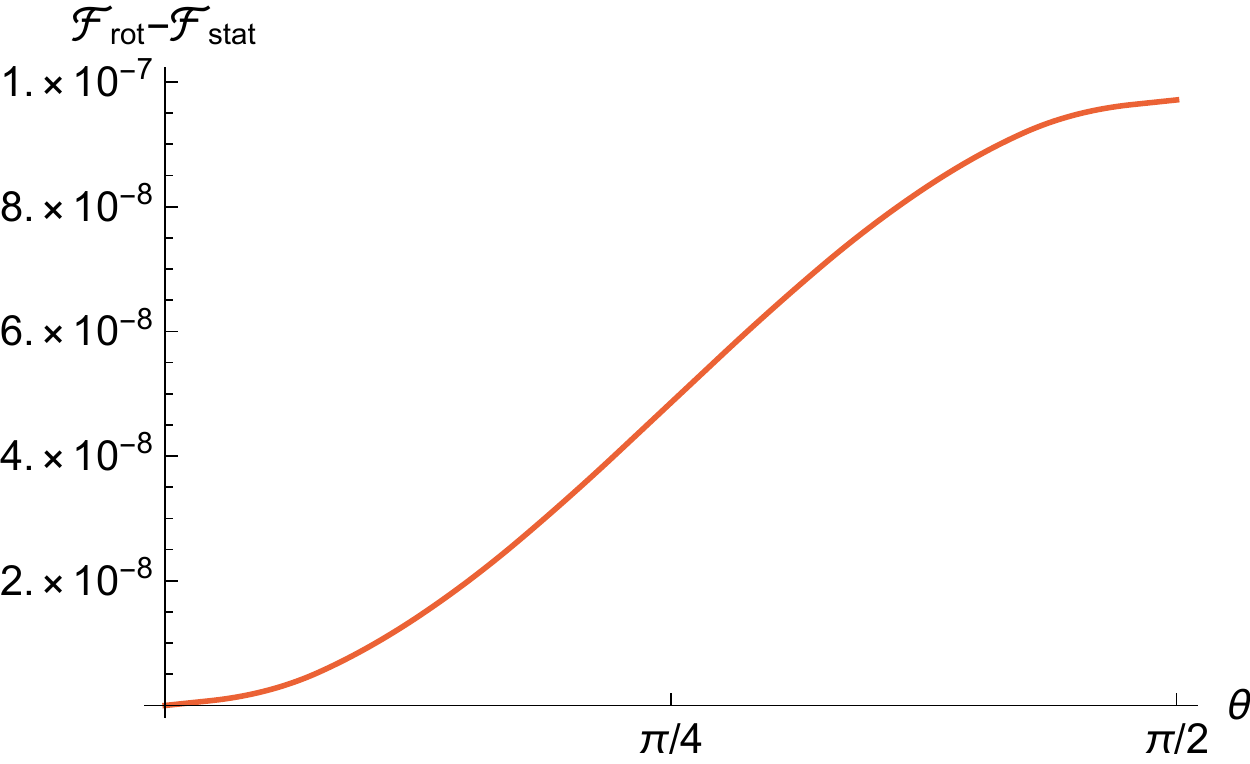}
\end{center}
\caption{
Detected difference between rotating and static shell depends of the distance of the detector from the center 
(left) and on the spherical latitudinal angular coordinate $\theta$ of the detector. Remaining parameters are described in Fig. \ref{fig:WanCongFig1}.
}
\label{fig:WanCongFig2}
\end{figure}

We show that the response function picks up the presence of rotation even though the spacetime inside the shell is flat and the detector is locally inertial. The detector can distinguish between the static situation when the shell is nonrotating and the stationary case when the shell rotates and the dragging of inertial frames, i.e. gravitomagnetic effects, arise. Moreover, it can do so when the detector is switched on for a finite time interval within which a light signal cannot travel to the shell and back to convey the presence of rotation. 

The summary of the results for quantum detection of the dragging of inertial frames is taken from 
the publication \cite{CBKM1}. 
(See also the contribution of W. Cong in the Session PT5 at MG 16.)

\section{Dragging effects of gravitational waves}
Rotating gravitational waves can also become a source of the dragging. 
The situation when the central frame is surrounded by rotating gravitational waves 
was for the first time modeled assuming
the translational symmetry along $z$-axis in \cite{BiKaDLB,BKL08cqg2}. 
Although this assumption implies unbounded energy of the gravitational waves and the spacetime is not asymptotically flat, the problem can be treated analytically as the master equation for the 
single function describing the gravitational wave has the form of a flat-space wave equation $\square \psi(t,\rho,\varphi)=0$.
Given a particular solution to this equation, other metric functions appearing in the line element 
\begin{equation}
 ds^2 = e^{2\gamma-2\psi}(dt^2-d\rho^2) -W^2 e^{-2\psi}(d\varphi+\omega\,dt)^2-e^{2\psi} dz^2
\end{equation}
can be determined from the Einstein equations. In particular, averaging of their $t-\varphi$ component
identifies $\langle\psi_{,t} \psi_{,\phi} \rangle $ as a source of the dragging of the inertial frames on the axis and for $\psi$ in the form of a cylindrical shell the central frame rotation is then found in a closed form. The analogy with angular momentum transport in spiral galaxies is discussed in \cite{BKL08cqg2}.

A similar problem permitting asymptotic flatness was then studied in 
\cite{BKLL2012,Barker2017CQG}. The gravitational waves are assumed to form a spherical shell described again by a single scalar function $\psi(t,\mathbf{x})$ satisfying flat-space wave equations $\square \psi = 0$ which this time appears only as the first order approximation of the full Einstein equations. The spacetime metric 
(in which we now use signature $-\!+\!+\!+$)
\begin{equation}
\label{eq:ds2Perturbed}
g_{\mu\nu} = \eta_{\mu\nu}+h^{(1)}_{\mu\nu}+h^{(2)}_{\mu\nu}+...
\end{equation}
is decomposed  into a flat Minkowski metric $\eta_{\mu\nu} = {\rm diag}(-1,1,r^2,r^2\sin^2\theta)$  in spherical coordinates $t, r, \theta, \varphi$, and the first and second-order perturbations $h^{(1)}_{\mu\nu}$ and $h^{(2)}_{\mu\nu}$. 
Then the first-order metric perturbations due to linearized gravitational waves appear the source of the second-order perturbations 
\begin{equation}
G_{\mu\nu}^{(1)}[{h^{(2)}}]= - G_{\mu\nu}^{(2)}[h^{(1)},h^{(1)}],
\label{eq:Gh2}
\end{equation}
where $G_{\mu\nu}^{(2)}[h^{(1)},h^{(1)}]$
contains terms of the Einstein tensor $G_{\mu\nu}$ quadratic in the first-order perturbations.

\begin{figure}[bp]
    \centering
    \includegraphics[width=0.81\linewidth]{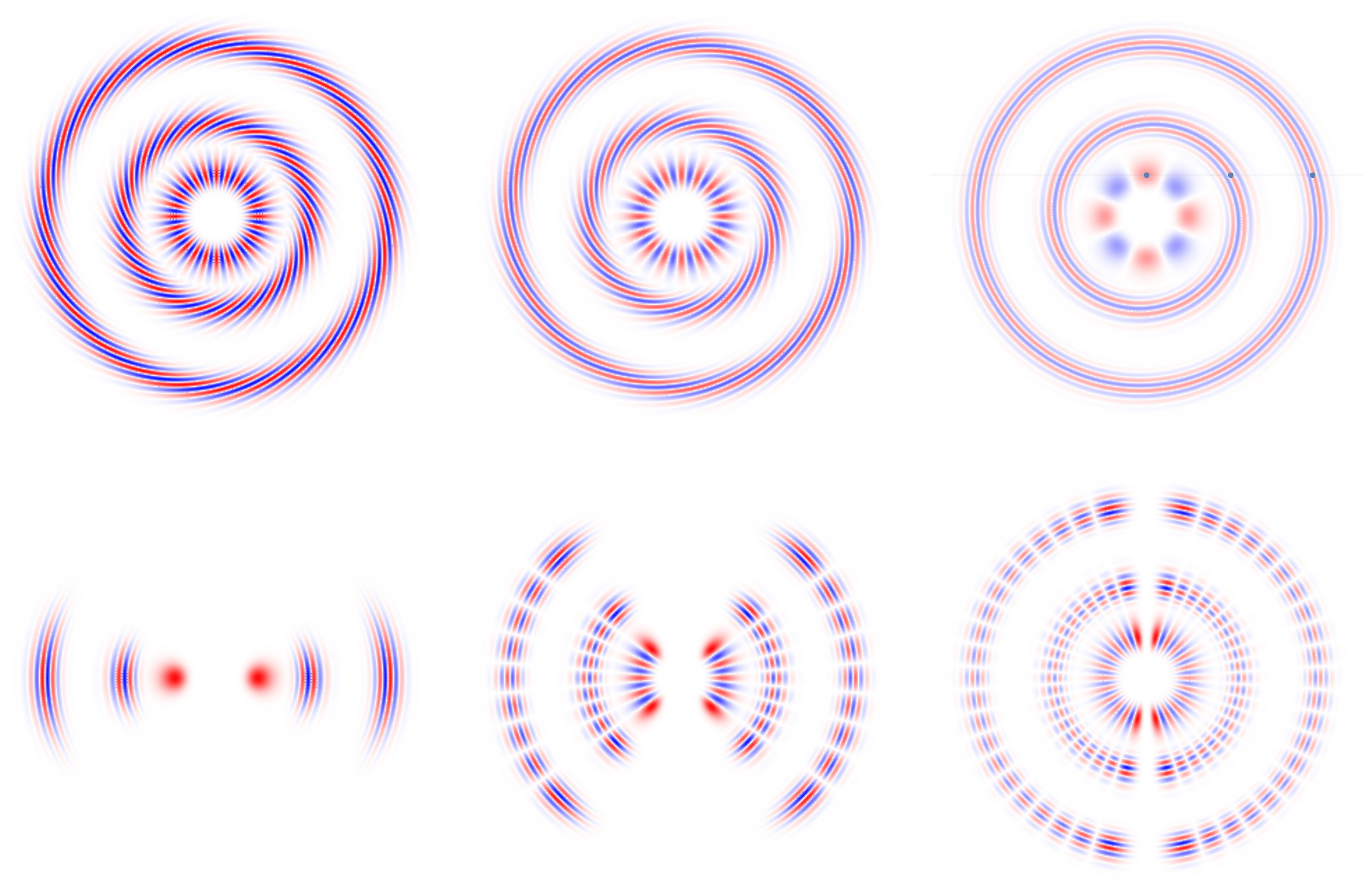}
    \\
    $l=24,m=24$~~~~~~~~~~~
    $l=24,m=16$~~~~~~~~~~~
    $l=24,m=4$
    \caption{Snapshots of the function $\psi$ in the equatorial plane $\theta=\pi/2$ (top) and in the meridional plane  $\varphi=0,\pi$ (bottom) at three distinct times $t=0,2a,4a$. The well-known behavior of spherical harmonics $Y_{lm}\sim \sin^{|m|}\theta$ means that for higher $m$
    the first order perturbations vanish not only near the center where we study the frame dragging but also along the $z$ axis.
    The top right plot also shows the position of a null particle with $\mathbf{r} = a\hat{\mathbf{y}}+t \hat{\mathbf{x}}$
    at given times to illustrate the localization of the wave at radii $r\approx \sqrt{a^2+t^2}$ (we denote Cartesian unit vectors $\hat{\mathbf{x}},\hat{\mathbf{y}}$, red/blue color indicates positive/negative $\psi$).
    }
    \label{fig:lmspheres}
\end{figure}

Assuming the Regge-Wheeler gauge, the function $\psi$ then directly determines 
$h^{(1)}_{t \theta},h^{(1)}_{t \varphi},h^{(1)}_{r \theta}$, and $h^{(1)}_{r \varphi}$, with remaining first-order perturbations vanishing, and the effects of the linearized gravitational waves are then 
determined by the analysis of the second-order terms.
To make space approximately flat for the central observer and his inertial frame, $h^{(1)}_{\mu\nu}$ is assumed to vanish near the origin.
Assuming a particular gauge, the quantity determining the central frame dragging can be determined from a quantity satisfying an elliptic equation, in a way similar to other situations. The central-frame rotation appears in the perturbation approach as a $l=1, m=0$ component in the expansion of 
 $h^{(2)}_{\mu\nu}$ into mutually orthogonal spherical tensor harmonics. Then a projection of Eq. \eqref{eq:Gh2} into the relevant $m=0$ tensor harmonic function resembles an averaging and yields  
 \begin{align}
\nonumber
&{{\frac{1}{2} }\!}\left[   {h_0^{(2)}}''\!\!\! - \frac{l(l+1)}{r^2} {h_0^{(2)}} -{{\dot h}_1^{(2)}}{}'\!-\frac{2}{r} {{\dot h}_1^{(2)}} \right]
 =\frac{1}{l(l+1)}\int_0^{2\pi}\int_0^\pi\!\!\!\!  G_{t\varphi }^{\,(2)}[h^{(1)} ,h^{(1)}]\partial_\theta Y_{l0}  \,\, d\theta\;d\varphi,
  \end{align}
where $h_0^{(2)}$ determines $h^{(2)}_{t \theta},h^{(2)}_{t \varphi}$,
and $h_1^{(2)}$ determines $h^{(2)}_{r \theta},h^{(2)}_{r \varphi}$ components 
of the second-order metric perturbations. Dots and primes denote the time and radial derivatives.
The rotation of the central frame $d\tilde\varphi=d\varphi-\omega_0 dt$ appears as $h^{(2)}_{t \varphi}=-\omega_0\,r^2\,\sin^2\theta$, so it is determined by the behavior of $h_0^{(2)}$  at $r=0$.
Using a global change of coordinate $\varphi\rightarrow \varphi+\delta\varphi^{(2)}(t,r)$, we can set 
\begin{equation}
h_1^{(2)}=0,~~~\text{i.e.},~~~h^{(2)}_{r\varphi}=0.    
\label{eq:h2gauge}
\end{equation}
Near the center we then have the Minkowski metric in spherical coordinates with the dominating perturbation corresponding to the slow rigid rotation of the central frame with angular velocity $\omega_0(t)$. Fixing the gauge condition \eqref{eq:h2gauge},
$h_{1\;l=1,m=0}^{(2)}=0$, prohibits any radial dependence of an additional coordinate transformation $\varphi\rightarrow \varphi+\delta\varphi^{(2)}(t,r)$ and the angle $\varphi$ in the center and thus also the central frame rotation $\omega_0$ is determined \emph{unambiguously} with respect to spatial infinity. We find
\begin{equation}
\omega_0=  \frac{1}{4 \pi} \iiint R_{t\varphi }^{(2)}[h^{(1)} ,h^{(1)}]~ \frac{\sin \theta}{r} dr\, d\theta\, d\varphi.
\label{eq:Omega0}
\end{equation}

\begin{figure}[tp]
    \centering
    \includegraphics[width=0.59\linewidth]{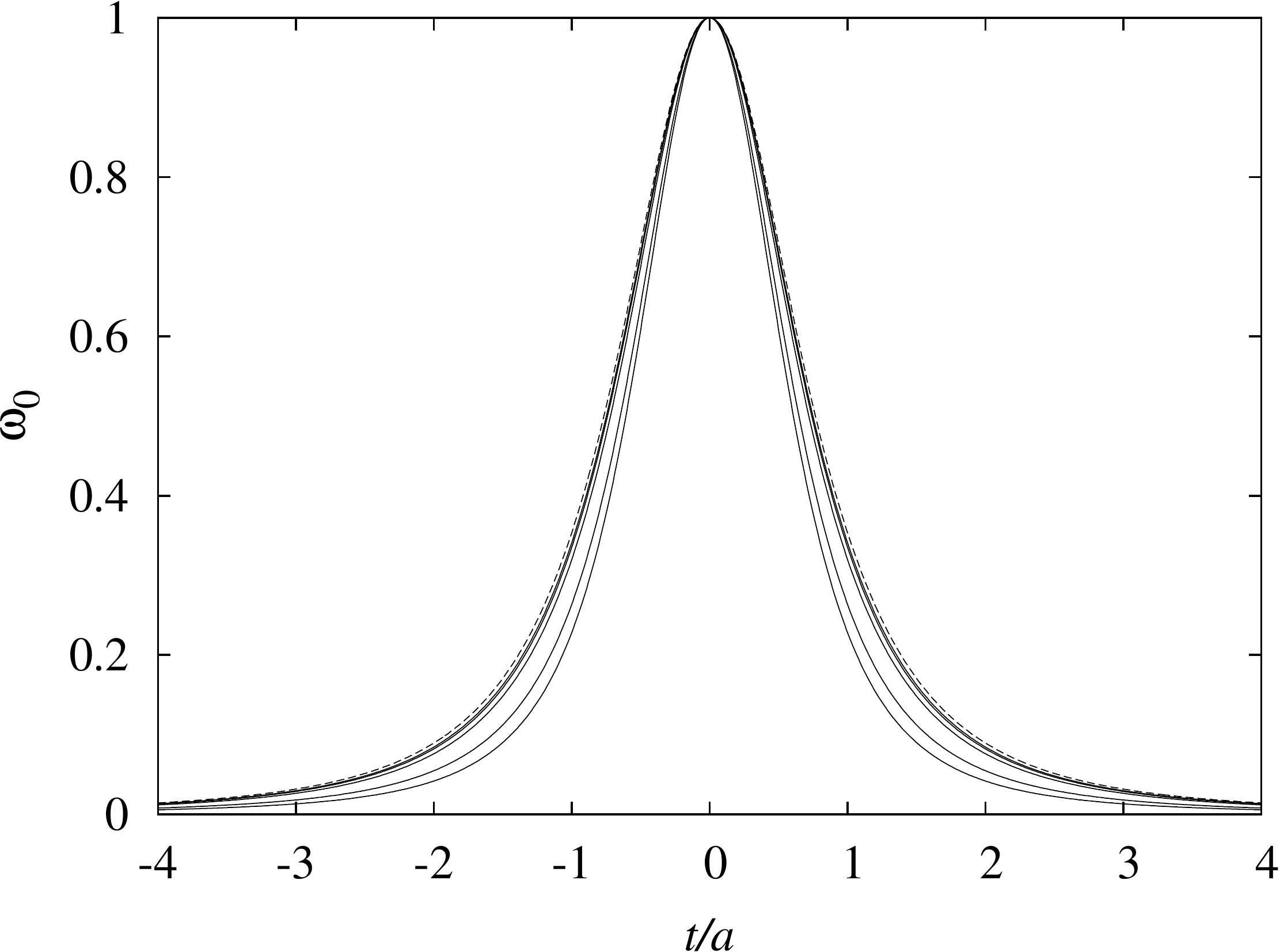}
    \caption{The dependence of the normalized angular velocity of the
central inertial frame $\omega_0(l, 1; t)/\omega_0(l, 1; t = 0)$ on the parameter $l = 2, 3, 10, 20, 30$ (from inside to out). The dependence \eqref{eq:omega0t} is shown as a dashed line.
    }
    \label{fig:omega0}
\end{figure}

To investigate further a particular closed-form solution $\psi$ of the wave equation, it has been chosen in the form of a shell of null radiation converging toward the origin, bouncing at the minimal radius $r \approx a$, and then expanding back to infinity (see Fig. \ref{fig:lmspheres}). This allowed us to evaluate the integral \eqref{eq:Omega0}
and find the explicit (though lengthy) formula for $\omega_0$. 
Assuming $l\gg 1$ it simplifies to
\begin{equation}
    \omega_0(t) \doteq \frac{\omega_0^{\text{max}}}{\left(1+\frac{t^2}{a^2}\right)^{3/2}}.
    \label{eq:omega0t}
\end{equation}
In the same limit, we show in \cite{Barker2017CQG} that the frame dragging is determined by the angular momentum of the gravitational wave $L_z$ and that the long exact formula can be approximated by $\omega_0^{\text{max}} \doteq 2L_z/a^3$. 
The angular momentum of the linearized gravitational waves is defined using the effective stress energy tensor 
\begin{equation}
L_z = - \int T^\text{eff}_{t\varphi} \;d^3 x,
~~~~
T^\text{eff}_{t\varphi} = \frac{1}{8\pi} G^{(2)}_{t\varphi}[h^{(1)},h^{(1)}].
\label{eq:LzGW}
\end{equation}
We can see that \eqref{eq:Omega0} and \eqref{eq:LzGW}
differ by a factor $r^3$ inside the integral. This explains why the approximate relation \eqref{eq:omega0t} holds: because for $l\gg 1$ function $\psi$ is localized around a thin shell with radius $r(t)\doteq \sqrt{a^2+t^2}$ (see Fig. \ref{fig:lmspheres}), the factor $r^3$ can be put in front of the integral. The time dependence of $\omega_0(t)$ on the parameter $l$
is shown in Fig. \ref{fig:omega0}.

\begin{figure}
    \centering  
 \includegraphics[width=0.34\linewidth]{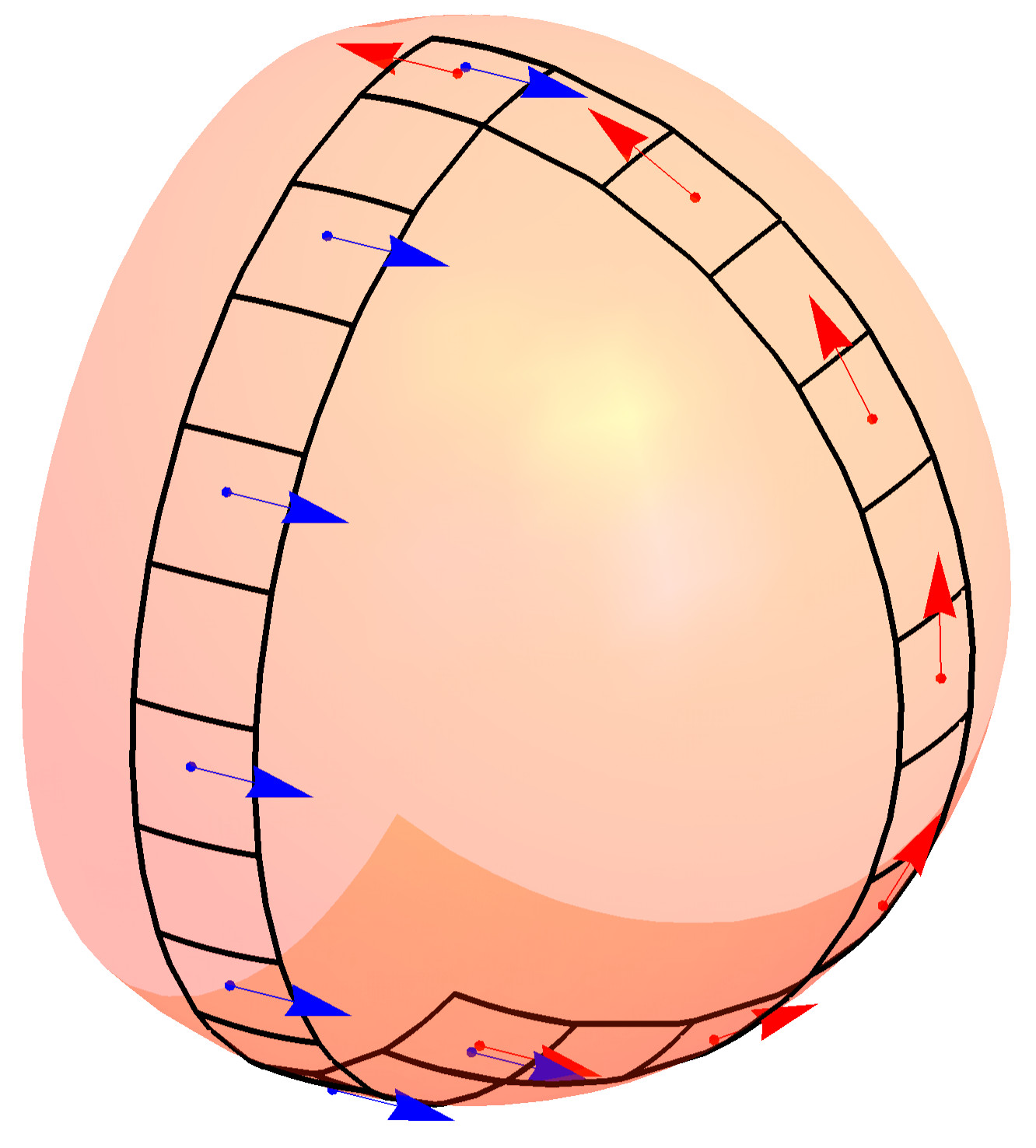}
 ~~~~~
 \begin{overpic}[permil,width=0.3\linewidth]{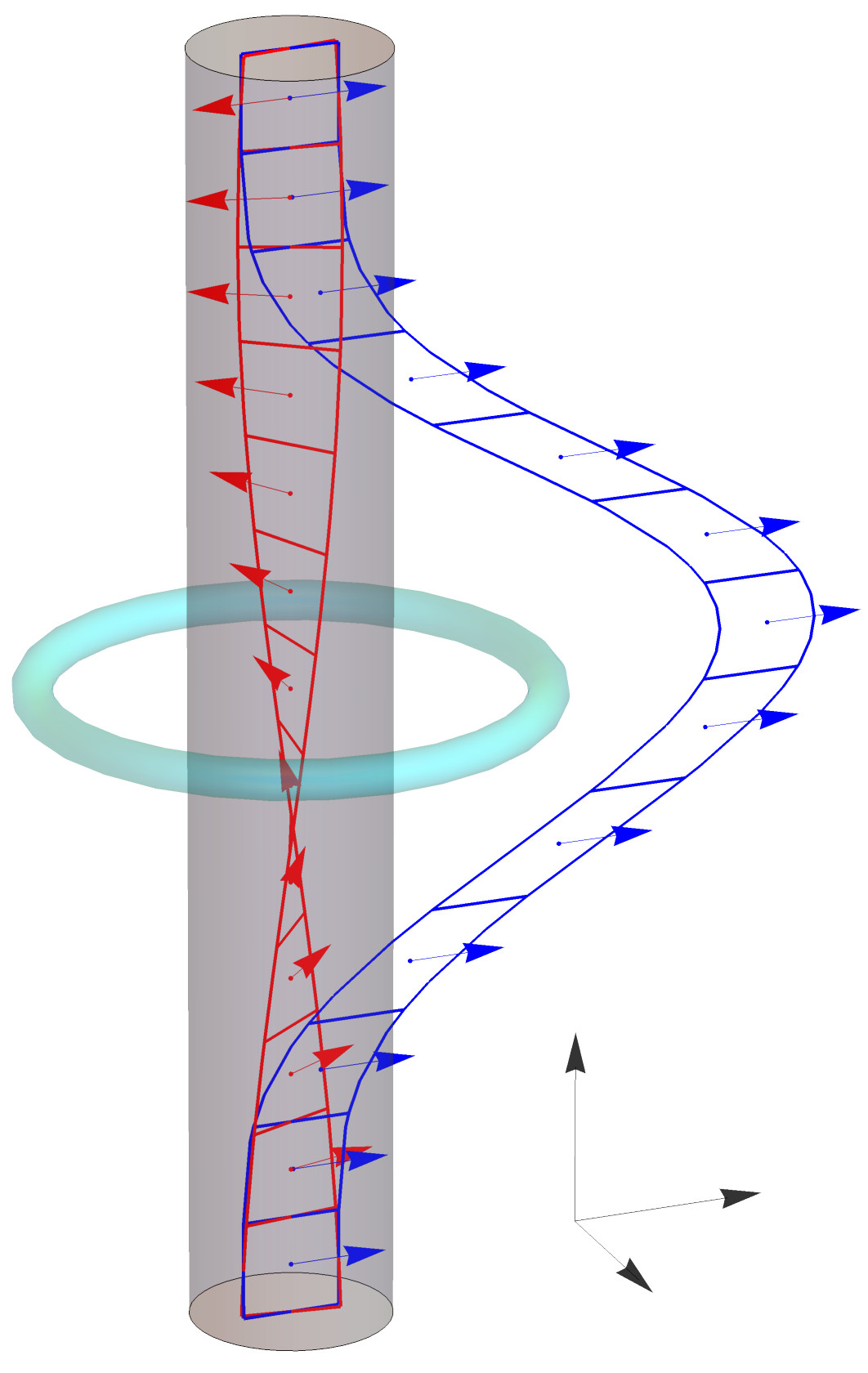}
  \put(410,60){\mbox{x}}
  \put(480,140){\mbox{y}}
  \put(390,170){\mbox{t}}
 \end{overpic}
    \caption{\emph{Left:} The fundamental dependence of parallel transport on the chosen path is usually demonstrated on a spherical surface naturally embedded in three-dimensional flat Euclidean space. Here we use neighborhoods of two meridians as an example of two approximately flat patches which yield mismatch when vector from the south pole is extended into both patches.  \emph{Right:} In our spacetime with rotating gravitational waves we also have two approximately flat patches. The spacetime is asymptotically flat which in the figure is symbolized by the blue ``ladder'' with arrows indicating ``fixed'' direction of a gyroscope. Because the gravitational waves do not reach the center, there is also approximately flat region near the center. Its worldtube is depicted as a gray cylinder. The gravitational waves are shown at the moment they are the strongest ($t=0$) as a blue torus encircling the central observer. The rotation of the central inertial frame (and gyroscopes there) is illustrated by the twist of the red spacetime-coordinate ``ladder'' and the gyroscope orientation.
    The mismatch of gyroscope directions at the top demonstrates the meaning of Eq. \eqref{eq:deltaphi0} as the implication of a particular form of spacetime curvature accompanying the rotating gravitational wave.
    }
    \label{fig:spherecurved}
\end{figure}

In an asymptotically flat spacetime we have two special flat-space worldlines categories --- the one of the central observer and that of a cautious observer who slowly retreats to $r\gg a$ so that she never meets significant metric perturbations. The discrepancy between the orientation of the gyroscopes following these worldlines 
\begin{equation}
    \Delta\varphi_0 = \int_{-\infty}^\infty \omega_0(t)\; dt 
    \label{eq:deltaphi0}
\end{equation}
can be seen as an illustration of the dependence of the parallel transport on the chosen worldline. 
In Fig. \ref{fig:spherecurved} we illustrate $\Delta\varphi_0$ as an obvious implication of the spacetime curvature due to the rotating gravitational waves. 
Thus, although the immediate value of $\omega_0$ involves instantaneous effects, its integral \eqref{eq:deltaphi0} representing the total rotation of the central gyroscope is a well-defined observable quantity. In the approximation $l\gg 1$ we then obtain $\Delta\varphi_0 \doteq 2a \omega_0^\text{max}\doteq 4L_z/a^2$. 
Such a simple relation is not available for dragging by a massive rotating shell, because its dynamics is not as unambiguous as that of gravitational radiation.

\section{On the dragging of inertial frames and Mach’s principle in cosmology}

In our treatment of the dragging of inertial frames in a cosmological context we shall mostly confine ourselves to the linear (cosmological) perturbation theory, rather than to exact models. Our inspiration will be Mach's principle as generally formulated by Hermann Bondi in his classical book \cite{Bondi1961}:
\emph{...all motions, velocities, rotations and accelerations are relative; local inertial frames are determined through distributions of energy and momentum in the Universe by some weighted averages of the apparent motions.}

 We started to realize such a ``Machian program'' in \cite{LKB}. We first analyzed frame-dragging effects due to slowly, rigidly rotating, but collapsing or expanding spheres in the (inhomogeneous) Lema\^{\i}tre-Tolman-Bondi universes, and we analysed the dragging effects of the vector perturbations of the FLRW universes described in a special gauge such that three (momentum) constraint equations enabled us to determine instantaneously metric perturbations $h_{0k}$ $(k = 1,2,3)$ in terms of energy-momentum perturbations $\delta T_{0k}$ and show how such averages are to be taken. In closed universes a linear combination of six Killing vectors (three rotations plus three quasi-translations) may be added to the $h_{0k}$. We also obtain the solutions of the three constraint equations when angular momenta corresponding to the three rotations and quasimomenta corresponding to the three quasitranslations of the sources (determined by $\delta T_{0k}$) are given. No absolute rotations exist in closed universe, only differences of rotation rates are determinable --- in accord with Mach’s ideas that `all motions are relative' (if the velocities of the bodies, described by perturbations of perfect fluid, are given, the metric perturbations are determined uniquely).
The last result is related to the fundamental fact that six globally conserved quantities, corresponding to the six Killing vectors in a FLRW universe, must all vanish if considered for the whole closed universe. 

It was, among others, an attempt to understand Mach’s principle in cosmological perturbation theory, which inspired us to formulate conservation laws even for large perturbations with respect to curved backgrounds \cite{KBL}. The resulting `KBL superpotential' (using the designation by Julia and Silva in their profound analysis \cite{JuliaSi}), was found, after applying certain natural criteria, to be unambigous and most satisfactory in spacetimes with or without a cosmological constant, in any spacetime dimension. It also found applications in various studies of generation of cosmological perturbations (see \cite{BKL} for references).
For the recent generalization to the Horndeski theory, see \cite{SchmidtBicak18}.

In a more recent paper \cite{BKL} we studied general linear perturbations of the FLRW universes from a `Machian perspective'. This led us to investigate both rotations and accelerations of local inertial frames in perturbed universes. 
We first introduced congruences of cosmological observers’ worldlines, defined their acceleration, rotation (twist, vorticity), shear and expansion in general, and then considered perturbed FLRW models
($g_{\mu\nu} = g^\text{FLRW}_{\mu\nu} + h_{\mu\nu}$). We found that un-acceler-\newline ated and non-rotating local inertial frames (LIFs) are determined by $h_{00,l}, h_{0l,m}, h_{0l,0}$.

We developed all the perturbed Einstein equations in a general gauge `ab initio', without assuming harmonic decompositions. Introducing the standard conformal time in FLRW universes, putting tildas over all the perturbation quantities, introduc-\newline ing the traceless part of $ h^l_k$ and notation

\begin{equation}
{\tilde {h}}{}^l_{T k}={\tilde{h}}^l_k
     -\tfrac13\delta^l_k{\tilde{h}}^n_n\ , ~~~~~
{\cal T}_k=\nabla_l {\tilde {h}}{}^l_{T k}
,~~~~~~
\mathcal{K} =\tfrac32 {\dot a}{\tilde h}_{00}\
         +\tfrac12 
         a \dot{\tilde h}^n_n
         -\nabla_l {\tilde h}^l_0,
\end{equation}
where $\nabla_l$ is the covariant derivative associated with the spatial FLRW background metric $f_{kl}$, $a$ is the expansion factor, dot the derivative w.r.t. standard cosmological time $t$ whereas the prime denotes the derivative with respect to conformal time $\eta, adt=d\eta$. Using $\nabla^2=f^{kl}\nabla_{kl}$, 
$k=0,\pm1$ for the curvature index and $\mathcal{H}=a H$, $H$ being the standard Hubble parameter, we find Einstein’s equations for perturbations to obtain the form
\begin{align}
    a^2\kappa\delta {\tilde T}^0_0=&a^2\delta {\tilde G}^0_0=
     \tfrac13 \nabla^2 {\tilde h}^n_n
     +k{\tilde{h}}^n_n-2{\cal{H}}\mathcal{K}
     -\tfrac12 \nabla_k {\cal T}^k,\\
a^2\kappa\delta {\tilde T}^0_k=&a^2\delta {\tilde G}^0_k=
     \tfrac12 \nabla^2{\tilde{h}}_{k0}
    +k{\tilde{h}}_{k0}
    +\tfrac16 \nabla_{kl}{\tilde{h}}^l_0
   {}+\tfrac23 \nabla_k \mathcal{K}
    -\tfrac12  \left({{\cal T}_k}\right)^{'},\\
a^2\kappa\left(\delta {\tilde T}^0_0-\delta {\tilde T}^n_n\right)=&
     a^2\left(\delta{\tilde G}^0_0-\delta {\tilde G}^n_n\right)=
     \nabla^2{\tilde h}_{00}
     {}+3a\left(\frac{1}{a}{\cal{H}}\right)^{\!'}{\tilde h}_{00}
     +\frac{2}{a}\left( a\mathcal{K}\right)^{'}, 
\end{align}
and
\begin{align}
a^2\kappa\left(\delta{\tilde T}^l_k-
     \tfrac13 \delta^l_k\delta {\tilde T}^n_n\right)=
     a^2\delta\tilde {G}{}^l_{T k} = ...~.
\end{align}
We do not write down fully the last equation since it describes waves and is not important for the determination of LIFs.  

To see how the LIFs can be determined by surrounding matter instantaneously on certain time-slices we use some specific gauges which we call the ``Machian gauges''. We give three examples of such gauges. For example, by putting $\mathcal{T}_k=0$ and $\mathcal{K}=0$, the first three equations become (hyper-) elliptic and the quantities determining LIFs can be found instantaneously when the (perturbations of) matter distribution are given.
The gauge conditions $\mathcal{T}_k=0$, fixing spatial coordinates, are associated with the ``transverse-traceless'' gauges in the linearized gravity and minimal-shear condition in numerical relativity. We assume these conditions to be valid in all three Machian gauges. In the first Machian gauge we choose the time slices to be so that $\mathcal{K}=0$. This implies the ``constant mean curvature slices'', and it coincides with Bardeen’s uniform-Hubble expansion gauge. In other two Machian gauges, together with the same gauge condition on spatial coordinates, we require ``uniform-intrinsic scalar curvature condition'' and the ``minimal-shear hypersurface condition'' (called the Poisson gauge by Bertschinger in 1995).  In \cite{BKL} these gauges are discussed in detail. In particular it is shown that they admit much less residual freedom than the synchronous gauge, frequently used in cosmology.

These Machian gauges have been considered in the group of D. Wiltshire, in particular, in \cite{Wilt1}, in the review \cite{Wilt2} and, most recently, by his students 
M. Williams \cite{MW} and R. Gaur \cite{G} in the context of the Post-Newtonian Cosmology.

We believe that dragging effects and Machian ideas will remain the source of inspiration.

\section*{Acknowledgments}
We acknowledge the partial support from Grant No. GAČR  21/11268S.

\end{document}